\documentclass[conference]{IEEEtran}
\IEEEoverridecommandlockouts

\usepackage{graphicx}
\usepackage{placeins}
\usepackage{url}
\usepackage{qtree}
\usepackage{titlesec}
\usepackage{eurosym}
\usepackage{multirow}
\usepackage{makecell}
\usepackage{array}
\usepackage{amsmath}
\usepackage{amssymb}
\usepackage{enumitem}
\usepackage{mdframed}
\usepackage{mathtools}
\usepackage{booktabs}
\def\BibTeX{{\rm B\kern-.05em{\sc i\kern-.025em b}\kern-.08em
    T\kern-.1667em\lower.7ex\hbox{E}\kern-.125emX}}
\begin{document}

\title{Neurosymbolic Architectural Reasoning: Towards Formal Analysis through Neural Software Architecture Inference}

\author{\IEEEauthorblockN{1\textsuperscript{st} Steffen Herbold}
\IEEEauthorblockA{\textit{Faculty of Computer Science and Mathematics} \\
\textit{Universität Passau}\\
Passau, Germany \\
Steffen.Herbold@uni-passau.de}
\and
\IEEEauthorblockN{2\textsuperscript{nd} Christoph Knieke}
\IEEEauthorblockA{\textit{Institute for Software and Systems Engineering} \\
\textit{Technische Universität Clausthal}\\
Clausthal-Zellerfeld, Germany \\
christoph.knieke@tu-clausthal.de}
\and
\IEEEauthorblockN{3\textsuperscript{rd} Andreas Rausch}
\IEEEauthorblockA{\textit{Institute for Software and Systems Engineering} \\
\textit{Technische Universität Clausthal}\\
Clausthal-Zellerfeld, Germany \\
andreas.rausch@tu-clausthal.de}
\and
\IEEEauthorblockN{4\textsuperscript{th} Christian Schindler}
\IEEEauthorblockA{\textit{Institute for Software and Systems Engineering} \\
\textit{Technische Universität Clausthal}\\
Clausthal-Zellerfeld, Germany \\
christian.schindler@tu-clausthal.de}
}

\maketitle

\begin{abstract}
Formal analysis to ensure adherence of software to defined architectural constraints is not yet broadly used within software development, due to the effort involved in defining formal architecture models. Within this paper, we outline neural architecture inference to solve the problem of having a formal architecture definition for subsequent symbolic reasoning over these architectures, enabling neurosymbolic architectural reasoning. We discuss how this approach works in general and outline a research agenda based on six general research question that need to be addressed, to achieve this vision. 
\end{abstract}

\begin{IEEEkeywords}
software architecture, neural networks, large language models, architecture inference, neurosymbolic architectural reasoning
\end{IEEEkeywords}

\section{Introduction}
Formal software architecture descriptions are a powerful tool that can be used to verify that software products fulfill certain requirements, e.g., to verify that the architecture is designed to prevent attacks or fault propagation based on being able to access sensitive information~\cite{CHONDAMRONGKUL2021102631}~\cite{6606612}. Nevertheless, due to the effort involved in creating a formal architecture description of any non-trivial system, the application of such techniques is typically restricted to safety critical components where such a verification is required, e.g., in the automotive domain~\cite{bahig2017formal}. This prevents the broader application of verification techniques based on formal architectures within software development and, thus, does not allow most software to benefit from the safety and security guarantees that reasoning based on formal architectures can provide. This hinders formal architectures from being an information source for tools that analyze code quality (e.g., PMD, SpotBugs). In an ideal world, the formal architecture of every project would be known in the same manner as control-flow graphs and dependency graphs are available and exploited in a similar manner by such tools and the formal methods building on top of such formal architectures would become commonplace. 

To be able to achieve this long-term vision, we first need foundational research that enables the creation of formal architectures with less effort, ideally automated. We believe that neural networks can help us to achieve this vision. On the one side, we have Large Language Models (LLMs), e.g., GPT-4~\cite{openai2023gpt4}, LLAMA~\cite{touvron2023llama}, CodeLLAMA~\cite{roziere2023code}, among many others. These models enable powerful new methods to deal with textual data, including code and natural language descriptions. Computer vision models like Segment Anything~\cite{kirillov2023segment} allow us to analyze visually depicted information, including diagrams about architecture. The multimodal capabilities of some LLMs like GPT-4 that enable the combined inference from text and images further increase the possibilities to deal with information about software architectures. 

Our core idea is that we believe neural networks and current formal reasoning methods can be integrated into what we refer to as \textit{neurosymbolic architectural reasoning}. The goal of this paper is to outline this concept and at the same time establish the open research challenges we need to deal with to transform this vision into practically feasible real-world use case. We start by discussing the challenges related to the inference of architectures from code and informal descriptions, including how to deal with possible differences between both information sources. Then, we proceed to discuss how inferred architectures that may contain mistakes due to imperfections in the inference process can, nevertheless, be used for formal reasoning. Throughout all these considerations, the soundness of the approaches will be a central challenge we consider, as this is at odds with the probabilistic nature of neural networks. As a result of our considerations, we raise six fundamental research questions that are further detailed by 18 subquestions that we believe guide a comprehensive research agenda towards achieving neurosymbolic architectural reasoning.

\section{Related work}

\subsection{Architecture inference from source code}
\label{sec:from-code}

Learning a software architecture from code involves the task of identifying and recovering the underlying architecture of a software system from its existing code base. The main objective of this process is to create a high-level architectural model, including its components, relationships, and the architectural patterns used in the implementation.

Static approaches focus on analyzing the source code of a software system without executing it \cite{schiewe2022advancing}. These techniques use various code analysis tools to extract information from the source code, such as parsing source code to extract syntax and control flow information or using metrics to quantify certain properties of the code. 
The main advantage of static analysis is that it can be performed relatively quickly and does not require executing the code. Additionally, static analysis is useful for identifying architectural patterns and relationships between components, as well as identifying potential issues in the source code~\cite{schiewe2022advancing}. However, static analysis has limitations, as it cannot account for the behavior of the software at runtime and may not reflect the actual architecture of the system~\cite{1631109}.

Dynamic approaches, on the other hand, focus on the actual runtime behavior of the system by analyzing the execution of the code. These techniques use dynamic analysis tools to capture and analyze the behavior of the system as it executes. By monitoring the system's execution, these approaches can identify the actual flow of control and interactions between components, enabling the identification of patterns that may not be visible in the static analysis of the source code \cite{ball1999concept}. Dynamic analysis is useful for detecting complex interactions between components and identifying performance and scalability issues in the system \cite{9063512}. However, dynamic analysis can be computationally expensive and may not scale well to large systems.

Schiewe et al. \cite{schiewe2022advancing} present a language agnostic approach to identify higher level components from source code. They introduce an intermediate level agnostic abstract syntax tree (LAAST) and identify component candidates based on heuristics.
Other approaches focus on the software architecture reconstruction of microservice applications \cite{gortney2022visualizing}. Walker et al. \cite{walker2021automatic} combine static analysis and dynamic tracing of service interactions to reconstruct the architecture of running microservice applications.


From the mining software repositories perspective, the static approaches discussed above can be seen as design structure, i.e., how the design and architecture is structurally reflected within the source code, e.g., through function call or similar relationships~\cite{maccormack2006}. An alternative perspective on architecture is through co-changes observed within software repositories, based on the assumption that files that are changed together are architecturally related~\cite{gall1998}. 
Co-changes indicate that components are strongly coupled, which in turn indicates a strong architectural relationship. By looking at how developers form teams that co-change files together, this provides a view on this sort of dependency, beyond the pure static structure. The importance of this additional perspective on software architecture is, e.g., demonstrated by recent work on the co-change of microservices. One of the core ideas behind microservices of architectures is that each service is a separated architectural unit that can be maintained independent of the others. However, research considering the co-changes of microservices instead shows that there are frequent co-changes beyond services boundaries~\cite{wesley2023}, including known code smells like shotgun surgery~\cite{fowler2018refactoring}. 

Recently, approaches from the field of machine learning in particular have been proposed to generate architecture descriptions from source code: 
Sajji et al. \cite{sajji2023methodology} 
introduce an automated approach for the creation of class diagrams from source code using Graph Neural Networks (GNNs).
Komolov et al. \cite{komolov2022towards}
propose a machine-learning-based approach for detecting software architecture, namely, MVP (Model–View–Presenter) and MVVM (Model–View–ViewModel) as target architectural patterns. They apply nine ML methods for detection of software architecture from source code metrics.
Puchala et al. \cite{puchala2022ensemble} propose an approach to architecture recovery using ensemble clustering and multiple dependencies. 
In the architecture recovery process structural, semantic, and directory dependencies from the software source code are utilized.

\subsection{Architecture inference from design documents}
\label{sec:from-docs}

Software documentation often contains informal descriptions of architecture, e.g., in form of natural language documentation, figures with main components, their organization in layer and/or their interactions. This sort of architecture definition is typically defined such that it can be easily created and understood by humans, but not by machines with automated processing due to the lack of formalism and the sheer number of different ways such a definition can be achieved. 

Several works address the task of generating formal descriptions from informal descriptions (such as requirements written in natural language). These works can be broadly categorized according to their degree of user intervention, the model they are trying to generate and the used approach. Table~\ref{tbl:heuristics} provides a brief overview of such works, with a focus on generating UML class diagrams. All of the referenced works have the usage of heuristic rules in common. Two surveys confirm that the utilization of diverse heuristic rules is a dominating trend in this research area~\cite{survey_1} \cite{survey_2}. An example for a reoccurring heuristic rule can be found in \cite{libya}. \textit{"C-Rule 1: Extract the common nouns (...), and proper nouns (...) from the text and map them to classes."} The role of Natural Language Processing (NLP) in these approaches is to the preprocessing of input texts, e.g., through Part of Speech Tagging (POS), Named Entity Recognition (NER), and dependency parsing with the goal to retrieve information that can be exploited by the aforementioned heuristic rules.

\begin{table}[!ht]
\centering
\begin{tabular}{llp{2.5cm}}
\toprule
\textbf{User Intervention} & \textbf{Approach} & \textbf{References} \\
\midrule
Without  & Heuristic rules & 
\cite{bajwa_2}, \cite{mohanan}, \cite{Ben}, \cite{Arora} \\ 
Without   & Heuristic rules + NLP & \cite{bajwa}, \cite{vinay} \\
With   & Heuristic rules & 
\cite{mova}, \cite{pope}, \cite{sharma} \\
With   & Heuristic rules + NLP & \cite{libya} \\ 
\bottomrule
\end{tabular}
\caption{Different variants of heuristic approaches for architecture inference and which have UML class diagrams as output.}
\label{tbl:heuristics}
\end{table}

Recently, studies have used neural network architectures on classifying images with UML diagrams according to their type but without focusing in depth on UML semantic elements:
The approach by Shcherban et. al. \cite{shcherban2021multiclass}, e.g., automatically identifies four most popular types of UML diagrams (class diagrams, use case diagrams, activity diagrams, and sequence diagrams) and non-UML diagrams from images by using five popular neural network architectures using transfer learning.

Only a few studies address semantic analysis of UML diagrams: Koenig et al. \cite{koenig2023neural}, e.g., propose the NEURAL-UML framework for training a learning model to categorize and locate semantic elements in class diagrams from an image. They focus on the task of detecting and recognizing classes and types of relationships in a UML class diagram.

\subsection{Alignment with intended architecture}
\label{sec:alignment}

To verify the conformance of implemented source code to high-level models of architectural design a variety of approaches from the field of Software Architecture Compliance Checking (SACC) were  proposed (e.g., \cite{pruijt2013architecture}). The objective of these approaches is to prevent or detect architectural erosion by comparison of the implemented architecture with the intended architecture. 
Further approaches provide repair recommendations (e.g., \cite{terra2015recommendation}) for cases in which deviations between both architectures are detected. 

Besides, there are approaches checking the consistency of the intended architecture against the selected reference architecture:
Bucaioni et al. \cite{bucaioni2023reference} present MORE, an approach that enables software architects to automatically check the compliance of their architecture description to previously modelled reference architectures.

\subsection{Formal reasoning with architectures}

A significant challenge in designing software-intensive systems involves architectural analysis, which entails identifying critical system properties through architectural models prior to implementation. 
Araujo et al. \cite{araujo2019research} evaluate the research landscape on formal verification of architecture descriptions. They identify what Architecture Description Languages (ADLs) have been used for architecture description towards supporting verification, their characteristics, and supported views.

A further direction towards formal reasoning with architectures are architectural tactics which are key abstraction of software architecture, and support the systematic design and analysis of software architectures to satisfy quality attributes. The work in \cite{marquez2023architectural} presents a systematic mapping study of architectural tactics in software architecture literature. 


\section{Approach and Challenges}

The starting point of our approach is the assumption that source code repositories and design documents already provide architecture definitions of the software. The source code can be seen as operationalization of the architecture. The design documents rather define the intended architecture, often on a more abstract level. Both are typically not directly suitable for formal reasoning over the architecture. Moreover, there are currently no guarantees that the source code actually follows what is prescribed by design documents. As a consequence, formal reasoning is not easily possible. Based on this starting point, we describe the challenges that our research community needs to address to overcome these issues and achieve neurosymbolic architectural reasoning and formulate concrete research questions that need to be addressed.

\subsection{Architecture inference from source code}

The source code of a product encodes the current architecture and can, therefore, be used as foundation to infer this architecture. Advances in neural networks through LLMs provide us with new and powerful methods to extract this architectural information from the source code. The drawback of neural methods is that they are unreliable, because of their probabilistic, black-box nature due to which we do not know their exact capabilities. This sort of inference is inductive and based on machine learning, i.e., correct with some likelihood and in flexibility constrained by the observed data. Recent related work started to consider such techniques (see Section~\ref{sec:from-code}). At the same time, there are already symbolic approaches based on static analysis that capture architectural considerations from source code, e.g., extracting components and their dependencies and the data flow between these components (see Section~\ref{sec:from-code}). These approaches are typically deductive, i.e., they are often defined based on hard-coded rules and we understand both the resulting correctness and the soundness.

For neurosymbolic architectural reasoning, we need to develop new methods that combine both worlds, i.e., the flexibility and power of the neural networks with the explainability and correctness of hard-coded rules. This requires us to find answers to the following open research questions:
\begin{itemize}[leftmargin=.34in]
    \item[\textbf{RQ1}] How sound and complete is neural architecture inference from code?
    \begin{itemize}[leftmargin=.27in]
        \item[\textbf{RQ1.1}] To which degree can specific architectural aspects (e.g., dependencies, layers, ...) be soundly inferred with neural networks from code, i.e., without risk of hallucinations?
        \item[\textbf{RQ1.2}] To which degree can specific architectural aspects be completely inferred with neural networks from code, i.e., without risk of missing architectural rules?
        \item[\textbf{RQ1.3}] Can neural architecture inference benefit from reasoning over multiple versions of code (i.e., its history) or is a single snapshot sufficient?
        \item[\textbf{RQ1.4}] How does the setting (e.g., programming language, size of project, architectural patterns used, ...) affect the soundness and completeness of the inference of architectures with LLMs?
        \item[\textbf{RQ1.5}] Do different formal paradigms for architecture definition affect the capabilities of the LLMs, i.e., does the syntax of the output affect the capabilities?
    \end{itemize}
    \item[\textbf{RQ2}] How can rule-based inference be combined with neural architecture inference?
    \begin{itemize}[leftmargin=.27in]
        \item[\textbf{RQ2.1}] Can LLMs be used to create sound symbolic rules for architecture inference to scale sound deductive architecture inference?
        \item[\textbf{RQ2.2}] Can sound symbolic rules be used for the finetuning of LLMs to improve their soundness when inferring architectural rules?
        \item[\textbf{RQ2.3}] Does the finetuning of LLMs based on sound symbolic approaches reduce the completeness with respect to architectural aspects that are not addressed by these rules, i.e., can the LLMs then still infer architectural aspects that could not be inferred with symbolic approaches?
        \item[\textbf{RQ2.4}] How can architectural aspects be inferred from LLMs be combined with those inferred with sound symbolic methods and what are the trade-offs for soundness and completeness?
    \end{itemize}
\end{itemize}

\subsection{Architecture inference from design documents}

New neural technologies not only increase our ability to reason about code, but also about natural language and images. These are the modes of information typically used within design documents to describe the intended architecture. Such design documents inform and guide both the initial development of a system, as well as the subsequent maintenance. Since these design documents are often informal through texts and informal diagrams, they are not within the focus of automated architecture inference so far, even though they should provide ground truth about architectures, assuming they are kept up-to-date. In some cases, semi-formal languages like UML are used to describe architectures. However, such languages are typically not used rigorously, i.e., they are rather used as graphical notation instead of as a formal model. Same as above, recent prior work provides a first look at the extraction of architectural information from such sources (see Section~\ref{sec:from-docs}).

We posit that incorporating this knowledge is a chance for neurosymbolic architectural reasoning, and needs to be considered by future research. However, same as for code, there are many, often similar, challenges. We note that we do not assume completeness to be as important as objective for inference from design documents, based on the assumptions that the design documents themselves are typically incomplete and only contain a high-level description of architectural constraints. Consequently, completeness cannot be expected and the problem rather shifts to ensuring soundness for the translation of the informally presented information to formal languages. 

\begin{itemize}[leftmargin=.34in]
    \item[\textbf{RQ3}] How sound is neural architecture inference from informal design documents?
    \begin{itemize}[leftmargin=.27in]
        \item[\textbf{RQ3.1}] How can we translate from the often ambiguous natural language to unambiguous formal architectural constraints without wrong assumptions about the meaning of language to achieve sound results?
        \item[\textbf{RQ3.2}] What are the limitations of current, powerful, computer vision approaches when it comes to the extraction of architectural constraints from images and how can these be overcome?
    \end{itemize}
\end{itemize}

\subsection{Alignment with intended architecture}

Since we are now dealing with two sources of architecture inference, the results may not agree with each other, i.e., what we inferred from code may not be the same as what we inferred from informal design documents. Unfortunately, understanding which is correct is non-trivial and likely cannot be fully automated. We possibly have unsound results in case architectural constrains where inferred with neural networks. Even if we have sound results, we do not know which is the correct architecture, because we cannot know if the developers deviate from a specified design intentionally, meaning the design documents are outdated, or whether this is unintentional, meaning that the source code violates the intended architecture. Especially the latter is a general concern, when we infer architecture from code without checking for its correctness: even if our inference is sound and complete, we might infer an architecture that does not reflect what the developers intend. We further note that while this sort of alignment is related to what was discussed in the literature (see Section~\ref{sec:alignment}), the key difference is that the literature assumes that the architecture definition is correct, which we cannot guarantee. 

All this raises additional challenges, that we believe can only be addressed by human oversight, i.e., by developers being in the loop, at least at some point, to ensure the inferred architecture used for formal reasoning aligns with the developer intention. 

\begin{itemize}[leftmargin=.34in]
    \item[\textbf{RQ4}] How can developers interact with neural architecture inference to validate results and improve soundness?
    \begin{itemize}[leftmargin=.27in]
        \item[\textbf{RQ4.1}] How can LLMs be used to infer architectures in such a way that they can subsequently be validated by humans in a short amount of time, thereby achieving soundness efficiently?
        \item[\textbf{RQ4.2}] How can we guarantee that symbolic rules for architecture inference created by LLMs can be validated by humans efficiently, i.e., such that their format and complexity do not make this prohibitively extensive?
        \item[\textbf{RQ4.3}] How can we present deviations between differences in inferred architectures to humans to enable the subsequent correction of mistakes and the alignment of architectures?
    \end{itemize} 
\end{itemize}

\subsection{Formal reasoning with inferred architectures}

The neural architecture inference is not a goal in itself, but rather only a requirement to achieve our goal, i.e., formal reasoning about these inferred architectures. Since the formal reasoning is deductive in its nature, soundness of the input is a premise that must be fulfilled. However, as outlined above, soundness is the key challenge when we infer architectures. Assuming that it will require a long time and that it may even be impossible with current technologies to achieve fully sound neural architecture inference, we need to adopt formal reasoning methods to be able to deal with, at least partially, unsound architectural inference. This requires us to either consider logics or other reasoning approaches that can deal with uncertainty or to deal with possibly flawed results of formal reasoning based on this broken premises. 

\begin{itemize}[leftmargin=.34in]
    \item[\textbf{RQ5}] How can we make formal reasoning about architectural constraints robust towards violations of the premise that the architecture definition is fully sound?
    \begin{itemize}[leftmargin=.27in]
        \item[\textbf{RQ5.1}] What are the consequences of mistakes in the architecture definition for the results of the formal reasoning, i.e., how is the soundness of the formal reasoning affected by the soundness of the architecture inference?
        \item[\textbf{RQ5.2}] What are alternative formal reasoning methods, e.g., based on fuzzy logics to deal with the uncertainty in the premises?
    \end{itemize} 
    \item[\textbf{RQ6}] What are the risks and limitations of using neural networks as part of a formal reasoning process?
    \begin{itemize}[leftmargin=.27in]
        \item[\textbf{RQ6.1}] How can we use neural networks to post-process the results of formal reasoning about architectures to mitigate possible mistakes due to wrong premises?
        \item[\textbf{RQ6.2}] How can we integrate neural networks directly into formal reasoning processes, such that formal reasoning is restricted to parts of the architecture that are known to be correct, while neural networks deal with uncertain parts?
    \end{itemize}
\end{itemize}

\section{Proof of Concept}

In our prior work, we already validated the feasibility of the general concept of neurosymbolic architecture inference and showed that we can leverage LLMs to learn a symbolic architecture rule~\cite{schindler2024formal}. Our primary goal was to derive a clear and concise software architecture rule (\textit{isAllowedToUse}) from an existing software system. We utilized the LLM to capture relationships and patterns in the software's structure. 

\begin{figure}[t]
    \centering
    \includegraphics[width=1.0\linewidth]{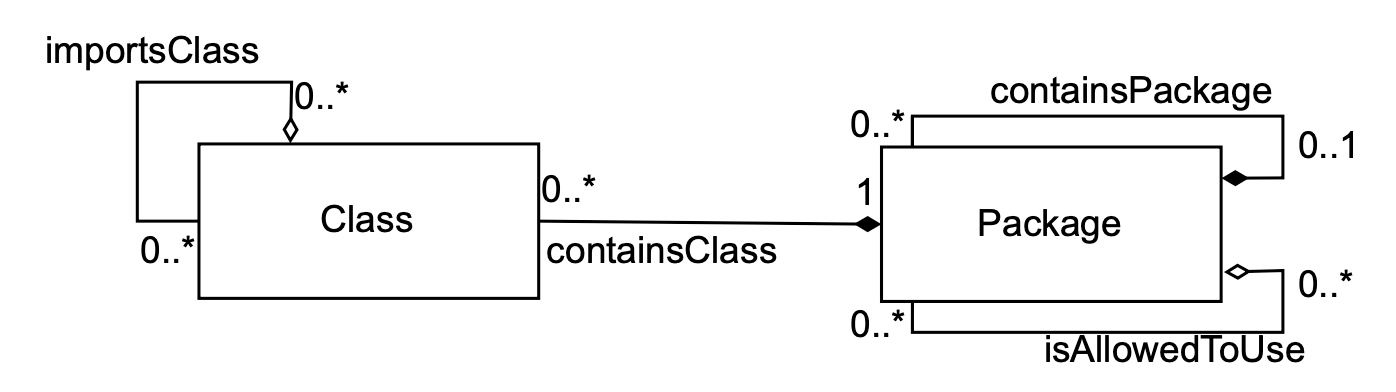}
    \caption{Domain model from \cite{schindler2024formal}}
    \label{fig:DomainModelExperiment}
\end{figure}

To achieve this, we first defined a domain model (see Fig.~\ref{fig:DomainModelExperiment}). The purpose of the domain model was (i) to facilitate the formulation of a reference rule and (ii) to enable the semantic mapping of source code into a formal knowledge base representation that matches this domain model. Structurally, the domain model encompasses relationships between classes and packages. Additionally, the domain knowledge facilitates the definition of rules that capture explicit connections between packages, thereby providing a structured basis for both rule formulation and code mapping.

As ground truth, we defined the following reference for the rule that should be inferred from source code examples: 
\begin{equation*}
\begin{split}
\forall a \forall b : isAllowedToUse(a,b) \Leftarrow package(a) \\ 
\land\ package(b) \land \exists c : class(c) 
\land \exists d : class(d) \\ 
\land\ containsClass(a, c) \land containsClass(b, d) \\ \land\ importsClass(c, d)
\end{split}
\end{equation*}

The experiments show that the LLMs were in principle able to detect the rule from positive examples, i.e., software that contained rules of this structure and also failed to identify such a rule for negative examples, as is expected. Thus, for the set of examples studied LLMs could be used in a strictly controlled setting as sound and complete method for the inference of architectural rules from code. Nevertheless, the results were not perfect, i.e., soundness could not be achieved for all examples. For example, there was a tendency to derive rules that are too simple and only closely resemble examples that were provided, hinting at limitations with the generalizability. 

When we consider this work in context of our research agenda, this is only a small aspect of \textbf{RQ1}. While we were able to demonstrate the general concept, this was only for a single, relatively simple architectural constraint that was tested in a rather controlled setting. Generalization to more rules, evaluation of soundness and completeness in other settings, or different formal languages for the formalization of the architectural rule were not yet considered. However, our example shows the incredible potential that modern neural networks have.

\section{Conclusion}

In this paper, we set out to define an ambitious research agenda for our community to achieve neurosymbolic architectural reasoning: the combination of neural networks with formal reasoning about software architecture, to enable such techniques broadly, ideally for every software project. We have showed that this is a complex endeavor that requires us to work on a large set of research problems, ranging from the inference of architectures from code and design documents with neural networks, over the dealing with the soundness and completeness of the results, to the impact of possible mistakes by neural networks on subsequent formal reasoning methods. We hope that this paper motivates not only us to continue working on this problem, but rather that the community comes together to address these major challenges together. 

\bibliographystyle{IEEEtran.bst}
\bibliography{literature}

\end{document}